\documentstyle[11pt,aaspp4,flushrt]{article} 
\begin{document}
\newcommand{\cmtwo}{cm$^{-2}$}
\newcommand{\degpoint}{\mbox{$^\circ\mskip-7.0mu.\,$}}
\newcommand{\kms}{\,km~s$^{-1}$}      
\newcommand{\minpoint}{\mbox{$'\mskip-4.7mu.\mskip0.8mu$}}
\newcommand{\mv}{\mbox{$m_{_V}$}}
\newcommand{\Mv}{\mbox{$M_{_V}$}}
\newcommand{\peryr}{\mbox{$\>\rm yr^{-1}$}}
\newcommand{\secpoint}{\mbox{$''\mskip-7.6mu.\,$}}
\newcommand{\sqdeg}{\mbox{${\rm deg}^2$}}
\newcommand{\squig}{\sim\!\!}
\newcommand{\subsun}{\mbox{$_{\odot}$}}
\newcommand{\et}{et al.~}
\newcommand{\cf}{c.f.~}
\newcommand{\eg}{e.g.~}

\def\ltsima{$\; \buildrel < \over \sim \;$}
\def\simlt{\lower.5ex\hbox{\ltsima}}
\def\gtsima{$\; \buildrel > \over \sim \;$}
\def\simgt{\lower.5ex\hbox{\gtsima}}
\def\arcs{$''~$}
\def\arcm{$'~$}

\title {THE DYNAMICS OF THE M87 GLOBULAR CLUSTER SYSTEM\altaffilmark{1}}
\author{\sc Judith G. Cohen\altaffilmark{2} and 
Anton Ryzhov\altaffilmark{2}\altaffilmark{3}}

\altaffiltext{1}{Based in large part on observations obtained at the
W.M. Keck Observatory, which is operated jointly by the California 
Institute of Technology and the University of California}
\altaffiltext{2}{Palomar Observatory, Mail Stop 105-24,
California Institute of Technology}
\altaffiltext{3}{Current address: Department of Physics, Box 351560,
University of Washington, Seattle, Washington 98195-1560}

\begin{abstract}
We present the results from a study of the dynamics of the system
of globular clusters around M87.  After eliminating foreground
galactic stars and background galaxies, we end up with a sample
of 205 bona fide M87 globular clusters for which we have radial
velocities determined from multi-slit spectra taken with the
LRIS on the Keck Telescope.
We find that the mean radial velocity of the M87 globular 
clusters agrees well with that of M87 itself, and that the
velocity histogram is well represented by a Gaussian distribution.
We find evidence for rotation in the globular cluster system.
We find that the observed velocity dispersion of the M87 globular cluster
system increases with radius from 270 \kms~ at $r$ = 9 kpc to
$\approx$400 \kms~ at $r$ = 40 kpc.  The inferred mass-to-light
ratio in solar units
increases from 5 at $r$ = 9 kpc to $\approx$30 at $r$ = 40 kpc
with $M(r) \sim r^{1.7}$.
The long slit optical spectroscopy near the center of M87 and
the recent analysis of the ROSAT X-ray data are in good agreement with
this analysis near the nucleus and in the outer parts of M87
respectively. 

\end{abstract}


\section{INTRODUCTION}

The existence of dark matter around spiral galaxies was established
by studies of their rotation curves at large radii using both
21 cm (Rogstad \& Shostak 1972) and optical (Rubin \et 1985)
emission lines from their gaseous disks.  As discussed in 
the review by de Zeeuw \& Franx (1991), for example,
elliptical galaxies in general
don't have much gas, and it has been difficult to find any
evidence for the possible presence of massive halos around them.

Globular clusters are ideal probes for studying the dynamics
of the outer parts of galaxies.  Their masses can be ignored
compared to the mass of the galaxy.  Except in the Galaxy, their
distances from the galactic center and their orbits cannot be  
determined directly, but they are plentiful enough around some galaxies
that statistics can be used to overcome this limitation.  One such case
is M87 at the center of the Virgo cluster,
which has an extremely rich globular cluster system, with 
thousands of members.  
4-m class telescopes are just large enough to reach the brightest
of these spectroscopically, and they
have been studied by
Mould, Oke \& Nemec (1987), Huchra \& Brodie (1987) and 
Mould \et (1990).  
Among the brightest objects, contamination by 
foreground galactic stars is
severe and this reduces the efficiency of isolating true M87 globular
clusters from among the candidates.  
 
Mould \et (1990) assembled the
data available at that time, combining the Palomar and MMT samples
to get a total sample of 43 globular clusters in M87.  Grillmair \et
have assembled a sample of similar size for globular clusters
in NGC 1399, the cD galaxy
in the Fornax cluster of galaxies.
But because of the small sample size, their estimates of the many critical 
kinematic parameters of these globular cluster systems, including
the rotation and the
velocity dispersion as a function of radius, are of low accuracy.
Merritt \& Tremblay (1993) estimated from numerical
simulations that a sample of $\approx$200
globular clusters in M87 is required to derive accurately the 
most fundamental parameter of the mass distribution of the halo
of M87,  the
exponent of the power law for the radial mass profile.

We present here a study of the kinematics of the globular clusters
around M87 with the goal of probing the distribution of mass around
this galaxy at the center of the Virgo cluster, deducing the
mass-to-light ratio in the halo of M87, and comparing our results
to inferences from recent X-ray missions
and from recent optical spectroscopy of the stellar component of M87.  
With the light
gathering power of the 10--m Keck I Telescope coupled with
an efficient multi-object spectrograph, we have succeeded in
overcoming the limitations of earlier work and have accumulated
a sample of more than 200 globular clusters in M87 with accurate
radial velocities.

A study of the abundances and ages of the M87 globular clusters that
can also be inferred from our data set will follow shortly.

\section{THE SAMPLE OF M87 GLOBULAR CLUSTERS}

The candidates were chosen from the photographic survey of Strom \et (1981)
for globular clusters in M87.  This survey 
of objects with $B < 23.5$ mag covers a square region
about 14 arc-min on a side.
The central region of M87 ($R < 60$ arc-sec) was excluded, 
presumably due to saturation of the photographic plates, and the
candidate list is incomplete out to $R < 90$ arc-sec.  There are 1728
objects in their sample.  Our sample includes
the 723 objects with $B < 22.5$ mag from the Strom \et survey.
We have found that their object positions
are adequate for designing slit masks that
align satisfactorily.  

A sample of bright globular cluster candidates near the center of
M87 was added to fill in the central region.
The radial distances from the center of M87 of these added objects
range from 24 to 71 arc-sec.  They were chosen from direct images
taken with the Low Resolution Imaging Spectrograph (LRIS, Oke \et 1995) 
on the 10--meter Keck I Telescope
and their positions relative to the bright globular cluster Strom 928, 
which is itself 23.2 arc-sec West and 62.3 arc-sec North of the nucleus
of M87, are given in Table~1.

Throughout this paper, 
we use the Strom \et (1981) running numbers to 
identify the clusters, while the 
additional objects near the center of M87 are denoted with running numbers 
beginning with 5001.

About half of the objects in the list of candidates
defined above were actually observed.  For each slitmask,
the first selection criterion was that the location of the candidate
be close to the centerline of the slitmask in the LRIS focal plane,
while the second one was
brightness.  The few 
very bright objects were avoided, as they are mostly galactic stars, while
objects in the faintest 0.5 mag of the sample were not included in 
slitmasks unless no other suitable candidate was available
near the desired position in the sky.

The spectra of about 15\% of the objects in these 14 slitmasks
were too noisy and no features could be recognized, hence
no radial velocity could be deduced.

14 M dwarfs (Strom 38, 87, 246, 354, 670, 681, 782, 801, 937, 996,
1549, 1551, 1608 and 1610)
and one white dwarf (Strom 178) were found.  These are
assumed to be galactic stars and are not considered further.

The sample of candidate globular clusters in M87 turns out to be
badly contaminated by galaxies, particularly in the outer
parts of M87.
Whenever possible, to eliminate background
galaxies, direct images taken with LRIS were used to verify that
candidate M87 globulars had 
images that were not extended.  Such LRIS direct images, 
particularly for the outermost parts of
M87, were not always available before the masks had to be designed.
Table~2 lists the 44 galaxies that were found.  If no redshift
is given, the galaxy was identified as such from LRIS direct images.

Figure 1 shows the distribution on the sky of the objects with measured
radial velocities, including the galactic stars
 and the galaxies, as well as that of the entire
$B < 22.5$ mag sample from Strom's lists.

%
%

%
%

\section{OBSERVATIONS AND DATA ANALYSIS}

We have used the LRIS
on the Keck I Telescope in the multi-slit mode.
Slitmasks whose length was 7.3 arc-min
were used to maximize the
efficiency of observations, with an average of 28 objects per
mask.  Two or three 3000
second exposures 
were obtained for each slitmask.  
All the observations used 600 g/mm gratings
(dispersion 1.24 \AA/pixel) with 1.4 arc-sec wide slits giving
an effective resolution of 8.1\AA~ (ignoring anamorphic magnification,
which improves this somewhat, especially in the red).  The
detector is a $2048 \times 2048$ pixel Tektronix CCD, so the
spectral coverage is $\approx$2500 \AA.

These observations were carried out on 7 nights over a 3 year period.
The seeing near the zenith  for these nights
ranged from 0.6 to 1.2 arc-sec.
  
A total of 14 different slitmasks have been used.  
In 1994, four masks were used
with the 600 g/mm grating blazed at 7500\AA ~centered at 8400\AA.
The dominant feature in these spectra is the infrared Ca triplet.  
In early 1994, the detector
electronics were noisier than they should have
been, a situation corrected in mid-1994. In 1995,
8 masks were used with that same grating centered at 
5500\AA ~so as to pick up the Mg triplet and nearby strong 
Fe blends, as well as the Balmer lines $H_{\alpha}$ and $H_{\beta}$.
These masks have a position
angle of 0$^{\circ}$ to 315$^{\circ}$ in 45$^{\circ}$ increments.
Each mask is laid out radially along its selected position angle,
with one end of the effective slit placed 
close to the nucleus of M87 near the inner radial limit
of the Strom \et (1981) survey.
In 1996, a new 600 g/mm grating became available with a bluer
blaze angle (5000\AA).  This was used with two
masks centered on the nucleus of M87.  The exposures here
were only 2500 sec long, with 2 or 3 exposures per slitmask.  
These spectra were
again centered at 5500\AA.  With the exception of the 1996 spectra,
all these observations used dual amplifier readouts to maximize the
efficiency of observing.  
Since these are multi-slit spectra,
the actual wavelength coverage for a particular object depends on its
position with respect to the center of the entrance field of the
LRIS.

The LRIS has some flexure.  In part because of this,
the spectra were not flat fielded, although they were corrected for the
difference in bias level and gain between the two amplifiers used 
in the dual amplifier readout.  After this, the cosmic rays
were removed. Then the individual slitlet spectra were cut out 
of the full CCD frame.  Next the two-dimensional spectrum of each slitlet
was corrected for
distortions both along and perpendicular to the dispersion, 
arising partially from distortions in the LRIS camera as well as
from the long slit length used by LRIS slitmasks.  
A second order polynomial was fit to
the sky across each pixel in the dispersion direction, the sky 
was removed, and the appropriate
sum along the slit yielded the spectrum for each object.  The
one-dimensional spectra from the
various exposures of a given object with a given mask were summed to
get the final result for each candidate.  

The dominant
problem in these spectra (at least after mid-1994, when the detector
electronics were fixed) is the non-uniformity of the width of the
slit.  The slits are punched by a numerically controlled machine, and
need to be carefully cleaned of chips prior to use.  Variations of
a few percent in width are common, variations of 5\% occasionally
occur, and even larger variations occur if the masks are not cleaned
properly.  Efforts are underway within CARA to reduce or 
eliminate this problem.
Correction of this variation of slit width is not straightforward given the
flexure, and was not incorporated into our analysis scripts.

To ensure the best possible accuracy, the wavelength calibration was
carried out using the most unblended night sky emission lines
within the spectra of the globular clusters themselves.
The region  of the infrared Ca triplet is full of such, and there the 
wavelength calibration is very good.
Typical fits (a third order polynomial suffices) including
14 night sky lines distributed along the length of the spectrum
have a 1$\sigma$ rms residual in the
fit of 0.2\AA~ (7 \kms).  The 5500\AA~ region has far fewer lines
both in the night sky spectrum (no Hg lines are seen from
Mauna Kea) and in the available arc spectra. 
The 5199\AA~ night sky emission line is the last detectable
night sky emission feature on the blue side of these spectra.
Its presence was crucial to anchor the wavelength scale.
When the background of the integrated stellar light of M87 
is very bright, this weak
emission line is sometimes lost.  In a very few cases, for $\lambda < 5577$\AA,
the wavelength scale at the blue end had to be determined from other
spectra with better arc coverage.
The wavelength scale of each exposure of each multislit 
was independently calibrated. 

All the data analysis was 
carried out using Figaro.

The spectra of the M87 globular clusters show no sign of spatial
or velocity resolution.  The ``sky'' spectra, consisting of 
the night sky and the halo light from M87 itself,
show obvious line broadening
for those globular clusters close to the center of M87,
presumably due to the velocity dispersion of the M87 stars.

\section{RADIAL VELOCITIES}

These spectra of candidate globular clusters in M87 
cover one of two different wavelength regions.
The radial velocities for the spectra centered at 8500\AA ~were obtained
by averaging those inferred from $H_{\alpha}$, which occasionally 
fell within the
wavelength regime covered, and
from each of the detected components of the infrared Ca triplet.
(Only for the brightest M87 globular clusters was the weakest of 
the triplet lines clearly detected.)  The radial velocities
for the spectra centered at 5500\AA ~were obtained via averaging the
results of cross
correlations covering the region from 5150 to 5400\AA~ and
also the $H_{\alpha}$ region.   $H_{\alpha}$ is outside the 
wavelength region covered
in about 25\% of these spectra.  While
$H_{\beta}$ is often included,
the wavelength scale at 4900\AA~ is poorly determined.  The brightest
M87 globular cluster in the particular mask being reduced
was used as a template for the spectra from that mask.  
Its radial velocity was determined
using the laboratory wavelengths of the Mg triplet, of the adjacent
strong Fe blends, and of $H_{\alpha}$.  The cross correlations
were checked by visual inspection of the spectra, as sometimes
spikes from cosmic rays throw the results off, and these had to be
removed manually from the spectra
or the radial velocity determined via eye estimates
of the line centers.

Table 3 lists the heliocentric radial velocity for 
229 candidate globular clusters in M87
in our sample.  The histogram of heliocentric radial velocities for the
candidate M87 globular clusters is shown in Figure 2.  It
is double peaked, with the majority of the objects
distributed about the radial velocity of M87, while about 10\%
of the objects reside in
a narrower peak centered on 0 \kms. Even though
the M dwarfs have already been excluded, substantial contamination
by galactic stars is still present.  Since it is impossible to
separate the galactic stars of spectral type K from M87 globular
clusters within this peak, we reject all objects with $v_r < 250$ \kms, 
almost all of which are undoubtedly galactic stars.  This brings the number
of objects in our sample down to 205.
To avoid skewing
the distributions, we must also reject objects in the corresponding interval
of the high velocity tail of the $v_r$ distribution.
However, there are no objects with such large positive 
radial velocities
in our sample.  

These 205 globular clusters in M87
with kinematic data from our LRIS observations are the sample
used in the remainder of this paper.

%
%
%

\subsection{Uncertainty of the Radial Velocities}

The accuracy of the radial velocities depends on 
many factors, including
the signal
in the spectrum of the object, the signal in the spectrum of
the background, the strength of the spectral features for each
object, the accuracy of the background subtraction algorithm used, etc.  
The signal in the spectrum of the object depends on the transparency
during the observations, the integration time (6000 or 9000 sec total),
and the accuracy of alignment of the object with the slitlet in the 
slitmask.  If the object is too close to the upper or lower edge
of the slitlet, sky subtraction becomes more difficult.

Another potential source of error is illumination effects.
In normal long slit spectra, the radial velocity for a point
source depends on where the source is located within the
width of the slit and on the seeing.  For multi-slit spectra, the
same effect operates.   But now it depends on the accuracy of the astrometry
and object positions used to design the slitmask and on the
accuracy of alignment of the objects on the sky to the physical slitmask
in the entrance aperture of the spectrograph.  The individual errors
may vary from object to object and cause a spurious velocity dispersion.

We denote the quantity of interest, the 1d difference between the location
of the object and the center of its slitlet as seen by the
spectrograph, both measured in the dispersion direction, as $\Delta(X)$.
Detailed logs of each slitmask alignment procedure are available for
most of the 1995 and 1996 LRIS observations.  They indicate
a median 1$\sigma$ rms value for $\Delta(X)$ of 0.4 arc-sec.
This corresponds to 2.3\AA~(135 \kms~ at 5100\AA, 81 \kms~ at 8500\AA).
Numerical simulations show that the actual errors introduced into the
data are at least a factor of two smaller than this due to the
smoothing effects of the seeing. 

These globular cluster
spectra have much higher signal-to-noise ratios than any previous
M87 spectra.  For a typical M87 globular cluster candidate
with $B = 21.5$ mag, we accumulate,
summed along the slit, $\approx$2220 detected
photoelectrons/pixel in the dispersion direction in the
sky-subtracted spectra
centered at 5500\AA~ per 3000 sec of exposure.  The sky, avoiding strong
emission lines, and in the outer parts of M87 is  
$<$10\% as bright.
In many cases the accuracy of the radial velocities are limited by
the systematics of the definition of the wavelength scale rather
than the signal-to-noise ratio of the spectra.  This subject will
be discussed in more detail in the second paper of this series
on the abundances in the M87 globular clusters.  Here we merely
assert that the 1$\sigma$ uncertainty in the radial velocities of Table 3
is under 100 \kms.

For high signal-to-noise spectra, one can fairly easily achieve
a centroiding
error of 0.1 pixels, which translates directly into a $v_r$ error. 
Our adopted $v_r$ error of 100 \kms~ translates into 1.4 pixels
(0.2 resolution elements) at 5100\AA~ and into 2.3 pixels
(0.35 resolution elements) at 8500 \AA. 
Thus our
adopted radial velocity error of 100 \kms~ is indeed very conservative.

There are several ways to demonstrate that our data
in fact achieve this level of accuracy.
First we note that the
mean heliocentric radial velocity for the 205 M87 globular clusters 
in our sample is 1301 $\pm$27 \kms.  The
mean heliocentric $v_r$ for M87 itself
based on long slit high dispersion spectroscopy of the nuclear region of the
galaxy is 1277 $\pm$5 \kms (van der Marel 1994).  The agreement is
extremely good.

Next we examine the objects in the final sample
which have been observed more than once.
42 objects were observed twice, while one was observed 3 times.
(Objects were observed twice because objects observed in the
Ca triplet region were not automatically eliminated when
designing masks to be used at 5500\AA.  In addition, 
in certain places on the sky there were not many suitable
objects, so the same candidate was selected to be included on 
multiple slitmasks.)
Figure 3 shows the histogram of the
difference between the two measurements for the
43 objects that have been observed twice (including one observed 3 times).
Table 4 gives the $\sigma$ for the velocity
difference between pairs of measurements where both spectra are in the
8500 \AA~region, pairs with 1 spectrum at the infrared Ca triplet
and the second at the Mg triplet, and pairs with both spectra
centered at the Mg triplet.  The final column of Table 4 gives
the $\sigma$ for a single radial velocity measurement inferred
from the $\sigma$ of the differences of pairs of measurements.
Table 4 confirms, as expected, that the accuracy of $v_r$ from the
spectra centered at 8500\AA~ is higher than that from the spectra
centered at 5100\AA.
Again, we conclude that 100 \kms~ is a conservative value for
the 1$\sigma$ uncertainties in our radial velocities of 
candidate globular clusters in M87. 

%
%
%

\section{ROTATION}

The isophotes of M87 are not circular.  Many analyses of surface
photometry for this galaxy exist (see, for example, Cohen 1986)
and they have established
that the ellipticity increases outwards, reaching about 10\% at
a semi-major axis of about 80 arc-sec, with a major axis position
angle of $155~{\pm}5^{\circ}$.

There is no evidence for rotation exceeding 20 \kms~ in amplitude
in M87 from optical spectroscopy.
However the available data is often restricted to a single
position angle, usually the minor axis, and to a small slit
length on the sky (an effort to obtain maximum spatial resolution
near the nucleus to look for a nuclear black hole).  Such is
the case for the kinematic studies
of Bender \et (1994) and van der Marel (1994).  
Jarvis \& Peletier (1991) looked at 4 position angles with a slit
that covered $\pm$25 arc-sec from the nucleus, and found a maximum
amplitude of rotation of 20 \kms.  The recent work of Sembach
\& Tonry (1996) reaches to about 100 arc-sec out from the nucleus of M87.
Although the authors say nothing about rotation, based on their
long slit spectrum along the major axis there is
marginal evidence for it with an amplitude of not more than
30 \kms~ and with the SE sector of the galaxy having the
larger radial velocity than the NW sector.

We solve for rotation in the globular cluster system of M87
using the Levenberg-Marquardt non-linear
least squares fitting routines from Press \et (1986)
to fit a sinusoid in position angle (plus a constant mean velocity)
to the observed radial velocities of the 205 globular clusters.
The results for all the clusters, and then for them divided into
two radial bins, are given in Table 5, where the
number of M87 globular clusters in the sample, the
mean $v_r$, the rotational velocity ($V_{rot}$), and position angle
at which the rotational contribution to $v_r$ is largest
are given.  The uncertainty in $V_{rot}$ is about 20\%.


Figure 4 shows the rotation curve as a function of position angle
for the 205 globular clusters over two cycles.  The solid line
indicates the fit to the sample as a whole given on 
the first line of Table 5.  This figure does not appear as convincing
as might be desired.  So we carried out simulations where the
velocities of the M87 globular clusters were kept
at their observed values, but their position
angles were replaced by random values between 0$^{\circ}$ and
360$^{\circ}$ and the data set was then analyzed for rotation.
We conducted 200 trials of this procedure.  These
trials indicate that only 4\% of the time does the artificial
data show rotation as large or larger than that of the actual
data for the M87 globular clusters presented here.

Thus we infer a definite rotation in M87 which is approximately
(to with 30$^{\circ}$) about
the minor axis of the galaxy as defined from its intensity isophotes.
$V_{rot}$ is about 100 \kms. This study,
Sembach \& Tonry's (1996) long slit spectroscopy, and Mould et al's (1990)
earlier analysis of the M87 globular cluster system all agree on
the direction of the rotation (larger
$v_r$ along the major axis to the SE, smaller $v_r$ to the NW).

%
%

We need to adopt a definite rotational velocity profile in order
to correct the observed radial velocities.  For this purpose we
adopt a truncated solid body law,
$V_{rot} = 0.80R~cos(\theta-PA_0)$ with the restriction
that $V_{rot} \le 180$ \kms, 
where $R$ is the projected
radius (in arc-sec) and $PA_0$ = 140$^{\circ}$.  The exact nature of
the form adopted for $V_{rot}$ is not important, since 
in the next section we will show that rotation is small
compared to the intrinsic velocity dispersion of the M87 globular
cluster system.

We have carried out a Kolmogorov-Smirnov test on the observed radial
velocity histogram of the M87 globular cluster system
with the rotational corrections included.  We find 
that more than 90\% of the samples of 205 points drawn from
a normal distribution with a mean and variance matching that
of our ensemble will differ more from a Gaussian than do the
actual data for the M87 globular clusters presented here.

%
%

\section{VELOCITY DISPERSIONS}

We are now in a position to compute the velocity dispersions.
The 205 objects believed to be bona fide globular clusters in M87
are included.
Heliocentric corrections are applied and the rotation
is removed using the prescription given above.

The M87 globular cluster sample is divided into a number of radial
bins, and the mean velocity and velocity dispersion are calculated
for each bin.  Then the most discrepant object is eliminated,
and the mean and dispersion are calculated again.
Table 6 gives the results with and without the rotational correction
described above.  The differences
between the two are not large.  The sampling uncertainties
in $\sigma(v_r)$ assuming $v_r$ has a Gaussian distribution
of $\sigma(v_r)$/$\sqrt{2 N_{bin}}$ are also given for one case.

In the final column of Table 6 we give the values to be adopted
for subsequent use.   They include the rotation correction, and
an instrumental uncertainty of 100 km/sec has been removed in 
quadrature.


The velocity dispersion for the early-type galaxies
in the Virgo cluster is 570 \kms~(Binggeli, Tammann \& Sandage 1987),
even larger than that of our outermost point.

%
%

\section {SIMPLE MASS MODELS}

Our goal is to establish in a definitive manner whether or not
dark matter exists in the halo of M87, how much dark matter
exists, and how certain we are of such claims.

We wish to find a distribution of mass within M87 that reproduces
the velocity dispersion as
a function of projected radius ($R$) that we have derived from
our observations of the M87 globular cluster system.  There are 
many observations of
the surface brightness of M87, so once a mass distribution is
available, the mass-to-light ratio can easily be computed.
``Light'' throughout this discussion refers to the integrated
light at $V$ (5500\AA) of M87, which is dominated by the
stellar component of M87, although the M87 globular clusters do
make a small contribution as well.

As a first reconnaissance, although we know that the outer isophotes 
of M87 are not
circular, the ellipticity is small, and we treat $\rho(r)$
as spherical where $r$ is the radius.
We begin by assuming that the mass is distributed as is the light of M87,
i.e. $M/L$ is constant.

Hernquist (1990) proposed a general model density distribution
for spherical galaxies which has some very desirable properties.
Over a wide range in radius the projected mass derived
from this density distribution matches the well known
$R^{0.25}$ dependence for surface brightness. 
Many properties of this density distribution, including
the line-of-sight velocity dispersion for both isotropic and 
circular orbits, can be expressed as analytical functions.
This model, which is the model used by Mould \et (1990),
has two free parameters, a spatial scale ($A$) and
a total mass, $M$.  

Since a Hernquist model with $A$ = 52 arc-sec (3.8 kpc)
fits the M87 surface brightness profile, we
proceed to use this analytical mass model. 
We use the surface brightness
measurements of Boroson, Thompson \& Shectman (1993)
and of de Vaucouleurs \& Nieto (1978) to
numerically integrate for the enclosed surface brightness ($L(R)$),
and use the model only to calculate the deprojection factor 
$L(r)/L(R)$.
We assume a galactic
absorption $A_V$ of 0.14 mag.  We assume a distance of M87
of 15 Mpc (corresponding to a scale on the sky of 73 pc/arc-sec)
based on the Cepheid results for spirals in the Virgo cluster 
(Saha \et 1995,
Pierce \et 1994, Ferrarese \et 1996).

We assume an isotropic system 
($\sigma^2(v_r) = \sigma^2(v_{\theta}) = \sigma^2(v_{\phi}))$.
We then adjust
the only remaining free parameter of the Hernquist (1990)
model, the total mass $M$, so that
the line-of-sight velocity dispersion matches the observed
values for the the first three radial bins.  The required
value of $M$ is $7.5 \times 10^{11} M\subsun$. 
The enclosed $M(r)$ relationship for this model, 
if represented as a power law, has an exponent of 0.5 over the region
of interest.  Equation 41 of
Hernquist (1990) is the analytical solution for the projected velocity
dispersion for isotropic orbits, which we use to
calculate the projected velocity dispersion we predict for the stars in M87.

The probes we are using, i.e. the
M87 globular clusters, have a more extended areal distribution,
$n_{GC}$, than does the M87 stellar halo light
(Harris 1986, Cohen 1986, McLaughlin, Harris \& Hanes, 1995).  $n_{GC}$
also has an approximately constant surface density in a core 60
arc-sec in radius (more than ten times larger than that of M87 itself)
(Lauer \& Kormendy 1986).  We adopt the analytical function fit
to $n_{GC}$ of
Merritt \& Tremblay (1993).  The Abel projection integral
is then numerically inverted to obtain the volume density
of globular clusters, $\rho_{GC}$.  

To calculate the projected velocity distribution expected
for the M87 globular clusters, the Jeans equation in spherical
coordinates (equation 4-55 of Binney \& Tremaine 1994)
was integrated numerically using the Hernquist potential
with the values for the parameters given above.

The solid curve in figure 5 shows the
predicted line-of-sight velocity dispersion as a function of
projected radius.  The thick solid line denotes that expected
for the M87 globular clusters, while the thin solid line denotes
that expected for the stellar halo of M87.  
The filled circles in the figure denote the
values deduced from our measurements of the
the M87 globular cluster system taken from the final column
of Table 6.
The results from optical long slit spectroscopy for
$R \ge 20$ arc-sec, half of
which are from
from Sargent \et (1978) and the remainder from
Bender \et (1994), van der Marel \et (1994), and
Sembach \& Tonry (1996), are indicated by the open circles.
(The two values adopted from Sembach \& Tonry (1996) have
been corrected downwards by 7\% as suggested in their paper.)
The agreement between the optical long slit spectroscopy and
our data is good. 

Figure 5 shows clearly how tightly confined
towards the nucleus the optical data is, an indication of the
intense interest in the search for a nuclear black hole as
well as the fact that the rapidly
decreasing surface brightness of the M87 stellar halo makes such 
observations at larger radii impossible.
 
The model with mass following the light of M87, which by definition
has a constant mass-to-light ratio, does not fit the data.

%
%

In an effort to find a suitable fit, we
try varying the orbital distribution of the M87 globular clusters.
The dot-dashed line in figure 5
represents the predicted line-of-sight velocity dispersion for
the M87 globular cluster system assuming
purely circular orbits in the Hernquist potential whose
parameters ($M$ and $A$) have the values
adopted above.   The dashed line
represents a system under these conditions with radial orbits.
This curve was obtained by numerical integration of the spherical
Jeans equation with the appropriate orbital anisotropy term.
One might adopt an orbital distribution for the M87 globlar
clusters that varies with $r$,
depending on one's prejudices about globular cluster formation and
survival.  For example, clusters with radial orbits might not survive
at small $r$, but might survive at large $r$ where they do
not undergo repeated passages through the nucleus of M87.
However, one expects a system with a fixed total mass and
with an intermediate family of orbits to lie within the area 
bounded by the upper and lower of the three thick curves in figure 5.  
The range of behavior shown by the various orbital distributions is not
capable of explaining the observed data within the assumption of
a fixed $M/L$ ratio.

The distribution
of $v_r$ of the M87 globular clusters provides an additional
constraint
on the orbital characteristics although it probably requires
a larger sample of M87 globular clusters to exploit this
(Merritt 1997, Merritt \et 1997).

To get some idea of what kind of a mass model is in fact required to
match the data, we adopt as a toy model a central point mass and
isotropic orbits for the probes.  The central mass required to match
the data in each radial bin,
$M(r)$, can then be calculated directly from 
${\sigma_v}^2 = GM/(4r)$ (equation 10-6 of Binney \& Tremaine 1994), 
where $r$ is a radius, not a projected
radius.  Assuming galaxies are located at their
projected radii, we calculate
$<{\sigma_v}^2><r>$ for each radial bin to obtain $M(r)$.
This yields the mass-to-light
ratios as a function of $r$ given in Table 7, with $M \sim r^{1.5}$.
The required $M/L$ in the outer part of M87 are
very large, more than 10 time those of the inner part of M87.

%

Changing the orbital characteristics 
for the M87 globular clusters (isotropic versus circular versus
radial orbits) and allowing for the corrections associated with a proper
calculation of the gravitating mass
will not alter the fundamental result that there must be
a substantial amount of dark matter in the outer part of M87,
that $M \sim R^{\alpha}$ with $\alpha > 1$.

\section{FINAL MASS MODEL}

To derive a mass model we solve the spherical Jeans equation
assuming
isotropic orbits and spherical
symmetry

$$ M(r) = -(r~\sigma(v)^2 /G) ~{\times}~ \lbrack 
{dln(\rho_{GC})\over{dln(r)}} + 
{dln\sigma(v)^2\over{dln(r)}}\rbrack{.} $$

To evaluate the first term on the right side we 
use $\rho_{GC}(r)$ calculated earlier.
The logarithmic derivative
here varies from $-$1.4 within the flat core of the globular
cluster distribution in the inner part of M87 to $-$3.6
for the outer part. 
We de-project the observed velocity dispersions
by fitting $\sigma(v)$ as a function of $R$ with a linear fit
($\sigma(v)$ (\kms) = 216 + 0.49 $R$(arc-sec)).
(The de-projection term for $\sigma(v)$
is not large, never exceeding the 15\%, and
numerical integrations using $\rho_{GC}(r)$
give very similar results.)
This allows us to evaluate the logarithmic derivative term
analytically -- it varies from $\approx$0.2 in the inner part of M87
to about 0.8 for the outer parts.

For our outermost point, 
$ {-dln(\rho_{GC})\over{dln(r)}} - 
{dln\sigma(v)^2\over{dln(r)}}$ is 2.7, while the toy model described
above has a factor of 4.  This indicates the magnitude of the 
correction factors from a simple central point mass spherical model
with isotropic velocities, and provides some guidance as
to the magnitude of the
remaining model dependent uncertainties, which are primarily the choice
of the orbital characteristics for the M87 globular cluster system.

The results are given in Table 8.  The uncertainties indicated for 
$M(r)$ are from sampling errors only and do not include
any contribution for modelling uncertainties, problems with
the rotation corrections, etc.  Also note that the $<M/L>$ at $r$ is the
ratio averaged over the enclosed volume, not the value at $r$ itself.

Sargent \et (1978) obtained $<M/L> {\approx}7$ (using
the same definition of ``light'' as we are using) between 
$r = 50$ and 71.8 arc-sec, their outermost point, a result
reproduced by Saglia, Bertin \& Stiavelli (1992) 
with their more sophisticated models.  
The agreement
for the region of overlap of the two data sets is good.

Mould \et (1990), with a very limited sample of data,
found $M(\rm{39 kpc}) = 2.4 \times 10^{12}M$\subsun, in reasonable agreement
with the above.  Our much larger sample has allowed us to put
their preliminary
results on a much firmer basis and to derive a detailed
model for the behavior of $M(r)$.

Fitting a power law to $M(r)$ yields an exponent of 1.7, and the
$M/L$ ratios must be very large ($\ge 30$) in the outer part of M87.  There
is no way of re-working the data, the assumptions, 
or the analysis that avoids this
conclusion.

\section{THE X-RAY VIEW OF M87}

Very extended  X-ray emission centered on M87 has been detected and
can also be used to infer the gravitating mass of M87.
Here one assumes hydrostatic equilibrium and that the velocity
distribution of the gas particles is Maxwellian (isotropic) with
a velocity dispersion related to the temperature $T_X$
of the X-ray emitting gas in the usual way.
The surface brightness of the X-ray emission can be measured
by an X-ray imaging camera with suitable sensitivity, and
reasonably precise measurements were obtained by the Einstein
satellite.  The problem is the determination of $T_X$ -- such
measurements were beyond the state of the art until quite recently.

Fabricant \& Gorenstein (1983) analyzed the Einstein data for M87.
They concluded that $M/L$ increased from $\approx$10 at 
$r$ = 1 arc-min
to over 180 at $r$ = 20 arc-min ($\approx$87 kpc).  They
found a total mass of
M87 within 20 arc-min of M87 of $1.2 - 1.9 \times 10^{13} M\subsun$.  Though
they could follow the surface brightness in a broad X-ray bandpass
out to 90 arc-min, they could not determine $T_X$ beyond 25 arc-min.
The uncertainty in their determinations of $T_X(r)$ was large,
and they quote results for many models of this parameter.  The values
given in Table 9 are for their model 1, an isothermal
model with $T_X$ = 3 keV.  
Tsai (1992, 1994) has tried a more complex multi-phase model of the
Einstein data. He obtains similar results.

Nulson \& B\"ohringer (1995) provide a determination of the mass of the
central Virgo Cluster using ROSAT data.  Their determinations for
$M(r)$ in the inner 100 kpc centered on M87 are about 2.5 times
smaller than those of Fabricant \& Gorenstein (1983), partly due
to their use of a lower $T_X$ there.  Meanwhile, ASCA, with
its superb spectral resolution in the X-ray regime, has observed
M87 as well (Matsumoto \et 1996).  The ASCA data convincingly
demonstrate that at least two components are required to
fit the X-ray spectra, with $T_X$ of 3 keV and 1.3 keV
respectively.  The hotter component has a more extended spatial
distribution, and for both components, $T_X$ is approximately
constant with $r$.  It is not obvious how to correct the
ROSAT $M(r)$ determinations in light of these new ASCA results,
as the ratio of the emission measure for the hot and cool components
of the gas is changing rapidly over the spatial region of interest.

The best power-law fit to the ROSAT results of Nulsen \& B\"ohringer (1995) 
in the regime $50 < r < 100$ kpc
is $M \sim r^{1.4}$, close to the $M \sim r^{1.7}$ found from
the M87 globular cluster system.

$M(r)$ for the outer part of
M87 is displayed in Figure 7.  The solid points indicate the
results from the present globular cluster data.  The
1$\sigma$ uncertainties shown include only
observational limitations and do not include the modeling
uncertainties, which are at most 50\%.  The best fit
power law to these data
is shown as the thick solid line.  The large open circle
is the result of Mould \et (1990), while the X-ray upper and lower
limits from Nulson \& B\"ohringer (1995) are shown as thin solid lines.
The small open circles denote the values of Sargent \et (1978)
from optical spectroscopy of the stellar component of M87.
The overall agreement between these various methods, which rely
on different sets of physical assumptions and measure the gravitating
mass using different components of M87 as probes, in the regions
of overlap is very gratifying.

%
\section{SUMMARY}

This paper presents an analysis of the dynamics of the M87 globular
cluster system with the goal of determining the $M/L$ ratio
in the outer parts of this massive galaxy in the core of the
Virgo cluster.
A sample of globular cluster candidates in M87 has been observed 
using slit-masks with the LRIS on the Keck I 10-meter telescope.
After eliminating foreground galactic stars and background galaxies,
we isolated 205 bona fide members of M87 globular cluster system.

These globular clusters belong to M87; their mean $v_r$ is that of M87
itself
to within the (very small) observational error.
Their velocity distribution is consistent with that of a Gaussian
about the mean $v_r$.
There is also evidence for rotation of the globular cluster system of M87.

The observed velocity dispersion of the M87 globular clusters
increases with projected radius from about 270 \kms~
at $r$ = 9 kpc
to $\approx 400$ \kms~ at $r = 40$ kpc.
The inferred $M/L$ increases from 5 for $r < 9$ kpc
to  ${\approx}30$ for $r \approx 30$ kpc, where ``light''
refers to the integrated light of the stellar component of M87
at 5500\AA.  $M$(44 kpc) = 
3 x 10$^{12}M\subsun$, with $M \sim r^{1.7}$ for 
$4 < R < 33$ kpc.
No permitted changes in the data analysis, assumptions, or model
can eliminate the need for a substantial extended halo of dark matter
in M87.

These results fit reasonably well with those of optical long
slit spectroscopy at the innermost radii and onto those from
X-ray satellites at the larger radii.

In spiral galaxies determination of the properties of the
dark matter halo from rotation curves
is complicated by the presence of multiple 
components (disk, bulge, and halo) (Kent 1987), and the properties
of at least the first two of these are strongly variable from galaxy 
to galaxy, as well as by
data which in most cases does not reach out as far in radius. 
In M87, however, the dark halo dominates at $r \ge 10$ kpc, and 
we have ben able to infer its spatial
distribution in some detail.  However M87 can hardly be described
as a normal elliptical galaxy since it is located at the center of
a massive cluster of galaxies surrounded by a very extensive
halo of hot gas.  An obvious task for the future is to look for
evidence of dark matter in and around normal elliptical galaxies.

A discussion of
the implications of these results for the formation of the globular
cluster system of M87 will be presented in paper II 
with the abundance analyses.

\acknowledgements
The entire Keck/LRIS user community owes a huge
debt to Jerry Nelson, Gerry Smith, Bev Oke, and many other people who
have worked to make the Keck Telescope and LRIS a reality.  We are grateful to
the W. M. Keck Foundation, and particularly its late president, Howard
Keck, for the vision to fund the construction of the W. M. Keck
Observatory.  We thank the referee, Tad Pryor, for his constructive
suggestions that improved the analysis presented here.

\newpage

%
%
\begin{deluxetable}{lrr|lrr}
\tablewidth{0pt}
\scriptsize
\tablecaption{Location of Additional Globular Clusters Near the Center of M87}
\tablehead{
   \colhead{ID Number}
 & \colhead{East \tablenotemark{a}}
 & \colhead{North \tablenotemark{a}}
 & \colhead{ID Number}
 & \colhead{East \tablenotemark{a}}
 & \colhead{North \tablenotemark{a}}
\\
   \colhead{}
 & \colhead{(arc-sec)}
 & \colhead{(arc-sec)}
 & \colhead{}
 & \colhead{(arc-sec)}
 & \colhead{(arc-sec)}
}
\startdata
 5001 & +47.8 & --99.1 &  5016 & +46.3  &  --42.0 \\ 
 5002 & +37.9 & --97.0 &  5017 & +50.5 &  --50.7 \\
 5003 & +24.2 & --96.1 &  5020 & --24.4  & --51.4 \\
 5005 & +34.7 & --85.7 &  5021 & +22.9 & --117.8 \\
 5008 & +51.6 & --75.8 &  5024 & --15.4 &  --8.4 \\
 5010 & --11.8 &  --55.9 & 5025 & --22.5 &  --61.9 \\
 5012 & +60.6  &  --99.1 &  5026 & --22.1 &  --7.3 \\
 5014 & +15.8  &  --24.6 &  5028 &  +64.5  &  --19.5 \\
 5015 & +34.3  & --21.8 
\enddata
\tablenotetext{a}{Offsets with respect to the M87 globular cluster Strom 928.}
\end{deluxetable}

%
%
\begin{deluxetable}{rr|rr|rr}
\tablewidth{0pc}
\scriptsize
\tablecaption{Galaxies Found in the Sample of M87 Globular Clusters}
\tablehead{
   \colhead{ID} & \colhead{$z$} & \colhead{ID} & \colhead{$z$} &
       \colhead{ID} & \colhead{$z$} \\
   \colhead{(Strom No.)} & \colhead{} &  \colhead{(Strom No.)} & \colhead{} & 
       \colhead{(Strom No.)} & \colhead{} 
  }
\startdata
 3 & 0.435 & 69 & 0.086 & 85 & 0.154 \\
 147 & 0.098 & 160 & 0.249 & 171 & 0.250 \\
 188 & 0.194 & 211 & 0.656 & 301 & 0.250 \\
 312 & 0.498 & 385 &  &  386 \\
 404 & 0.093 & 412 &  &  435 \\
 465 & 0.683 & 474 & & 481\tablenotemark{a} \\
 563 &  &  584 & 0.093 & 771 & 0.433 \\
 854 &  &  867 & 0.260 & 876 & 0.35~ \\
 942 &  &  943 &  &  944\tablenotemark{b} \\
 971 &  & 982 &  &  987 \\
 1021 & 0.078 & 1043 &  &  1061 & 0.163 \\
 1081\tablenotemark{b} &  &  1137 & 0.258 & 1221 \\
 1253 &  &  1256 &  &  1294 \\
 1338 &  &  1398 & 0.311 & 1440 & 0.262 \\
 1502 &  &  1591 & 0.447
\enddata
\tablenotetext{a}{The spectrum of this object shows broad emission lines.}
\tablenotetext{b}{This object consists of a galaxy with a nearby stellar object.}
\end{deluxetable}

%
%
\begin{deluxetable}{rr|rr|rr|rr|rr}
\tablewidth{0pt}
\scriptsize
\tablecaption{Radial Velocities for Candidate Globular Clusters in M87}
\tablehead{
   \colhead{ID\tablenotemark{a}}
 & \colhead{$v_r$\tablenotemark{b}}
 &  \colhead{ID\tablenotemark{a}}
 & \colhead{$v_r$\tablenotemark{b}}
 &  \colhead{ID\tablenotemark{a}}
 & \colhead{$v_r$\tablenotemark{b}}
 &  \colhead{ID\tablenotemark{a}}
 & \colhead{$v_r$\tablenotemark{b}}
 &  \colhead{ID\tablenotemark{a}}
 & \colhead{$v_r$\tablenotemark{b}} 
\\
    \colhead{}
 &  \colhead{(\kms)}
 &  \colhead{}
 &  \colhead{(\kms)}
 &  \colhead{}
 &  \colhead{(\kms)}
  & \colhead{}
 &  \colhead{(\kms)}
  & \colhead{}
 &  \colhead{(\kms)}
 }
\startdata
 5021 &  1767 &  5002 & 1252 &  5005 &  1178 &  5008 &   195 &  5010 &  1401 \\ 
 5016 &  1478 &  5014 &  1278 &  5024 & 1076 &  5026 &  1990 &  5020 &  1646 \\ 
 5025 &  1762 &  5015 &  1669 & 5028 & 1414 &  5017 &  1173 &   5003 &  1294 \\ 
  5001 &   929 &  5012 &  1718 &  58 &  1923 &   59 &   $-$28 &   66 &  2260 \\ 
  91 &   179 &  101 &  1332 &  107 &  1515 &  141 &  1115 &  176 &  2210 \\ 
 177 &  1629 &  186 &  1761 &  191 &   716 &  235 &    31 &  248 &  1016 \\ 
 252 &   $-$41 &  279 &   820 &  280 &  1044 &  286 &    12 &  290 &  1413 \\ 
 292 &   847 &  307 &  1219 &  311 &   781 &  313 &  1768 &  314 &  1236 \\ 
 321 &  1376 &  323 &  1124 &  324 &   359 &  348 &   817 &  330 &   736 \\ 
 350 &  1244 &  357 &   $-$86 &  376 &  1182 &  378 &  1978 &  395 &  1890 \\ 
 410 &    15 &  417 &  1910 &  418 &  1866 &  420 &  $-$153 &  421 &  1724 \\ 
 423 &  1081 &  442 &  1424 &  453 &  1980 &  490 &  1570 &  491 &  1069 \\ 
 492 &  1498 &  518 &    42 &  519 &  1315 &  526 &  1169 &  537 &  1467 \\ 
 571 &  1757 &  579 &   999 &  581 &  1507 &  588 &  1637 &  602 &   555 \\ 
 611 &  1319 &  614 &  1891 &  645 &  1760 &  647 &   972 &  649 &  1373 \\ 
 651 &  2110 &  664 &  1563 &  672 &   702 &  678 &   $-$94 &  679 &  1218 \\ 
 680 &  1793 &  686 &   784 &  695 &  1869 &  697 &  1203 &  714 &  1315 \\ 
 715 &   467 &  723 &  1365 &  741 &  1212 &  746 &  1266 &  750 &  1406 \\ 
 770 &  1481 &  784 &  1891 &  787 &  1097 &  796 &  1130 &  798 &   950 \\ 
 801 &   $-$30 &  809 &   612 &  811 &  1436 &  814 &  1331 &  824 &  1212 \\ 
 825 &  1109 &  827 &  1477 &  831 &  1148 &  838 &  1086 &  849 &  1650 \\ 
 857 &   759 &  868 &  1480 &  871 &  1073 &  881 &   796 &  887 &  1778 \\ 
 892 &  1879 &  902 &  1587 &  904 &   914 &  910 &  1007 &  917 &   882 \\ 
 922 &  1760 &  928 &  1327 &  937 &   930 &  941 &  1140 &  946 &  1131 \\ 
 947 &  1519 &  952 &  1454 &  965 &  1362 &  968 &  1091 &  970 &   991 \\ 
 973 &   172 &  978 &  1878 & \tablenotemark{c} &  1141 &  991 &   950 &  992 &   727 \\ 
1007 &  1298 & 1010 &  1507 & 1015 &  1165 & 1016 &  1375 & 1019 &   650 \\ 
1023 &  1164 & 1032 &  1684 & 1034 &  1299 & 1044 &  1947 & 1049 &  1569 \\ 
1055 &  1543 & 1060 &  1622 & 1064 &  1405 & 1065 &  1100 & 1067 &  1368 \\ 
1070 &  1437 & 1091 &  1015 & 1093 &   905 & 1101 &  1494 & 1103 &   $-$31 \\ 
1108 &  1930 & 1110 &  1091 & 1113 &  1544 & 1116 &  1084 & 1119 &  1997 \\ 
1120 &  1367 & 1144 &   808 & 1155 &  1370 & 1157 &  1731 & 1158 &  1048 \\ 
1165 &  1473 & 1167 &  1476 & 1180 &  1106 & 1181 &   639 & 1200 &   878 \\ 
1201 &  1178 & 1205 &  1011 & 1208 &  1605 & 1216 &   101 & 1217 &  1125 \\ 
1219 &  1244 & 1220 &   855 & 1238 &   727 & 1240 &  1359 & 1244 &  1913 \\ 
1247 &  1672 & 1254 &   877 & 1264 &    39 & 1280 &     0 & 1290 &   728 \\ 
1291 &   $-$57 & 1293 &   776 & 1298 &   195 & 1301 &  1053 & 1309 &   728 \\ 
1313 &  1264 & 1322 &  1333 & 1336 &   969 & 1340 &  1296 & 1341 &   $-$41 \\ 
1344 &   982 & 1346 &  $-$178 & 1351 &  1705 & 1353 &  2161 & 1356 &     3 \\ 
1367 &  1278 & 1370 &  1041 & 1382 &  1472 & 1391 &  1186 & 1400 &   915 \\ 
1409 &  1103 & 1425 &  1338 & 1431 &  1303 & 1433 &  1917 & 1449 &  1067 \\ 
1457 &   784 & 1461 &   693 & 1463 &  1871 & 1469 &  1101 & 1479 &   447 \\ 
1481 &  1789 & 1483 &  1625 & 1490 &  1403 & 1497 &     3 & 1504 &   626 \\ 
1514 &  1165 & 1531 &   209 & 1538 &  1237 & 1540 &   587 & 1548 &  1735 \\ 
1563 &   695 & 1565 &  1574 & 1577 &  1366 & 1594 &  1537 & 1615 &  1168 \\ 
1617 &  1369 & 1631 &  1157 & 1664 &   995 & 1709 &  1790   
\enddata
\tablenotetext{a}{Identification from Strom \et (1981) except
for added clusters near center of M87 -- see Table 1.}
\tablenotetext{b}{Heliocentric radial velocities.}
\tablenotetext{c}{The spectrum of a second object 6.6 arc-sec E 
of Strom 978 was found in the slitlet of Strom 978.}
\end{deluxetable}

\newpage
%
%
\begin{deluxetable}{crrr}
\tablewidth{0pc}
\scriptsize
\tablecaption{$v_r$ Comparison for Objects Observed Twice}
\tablehead{
\colhead{Spectral Range\tablenotemark{a}} & \colhead{No. of Objects} & 
\colhead{$\sigma[\Delta(v)]$} &
\colhead{Inferred $\sigma(v_r)$} \nl
\colhead{(\AA)} & \colhead{} & \colhead{(\kms)} & \colhead{(\kms)}  }
\startdata
Any  &  43 &  117 &  83 \nl
5200 -- 5200 & 17 & 134 & 95 \nl
5200 -- 8500 & 23 & 80 & 68\tablenotemark{b} \nl
8500 -- 8500 & 3 & 60 & 42 \nl
\enddata
\tablenotetext{a}{Central wavelength of multi-slit spectrum.}
\tablenotetext{b}{This applies to a spectrum at 5100\AA~ assuming the results
of the last line of this table.}
\end{deluxetable}

\clearpage
\newpage
%
%
\begin{deluxetable}{crrrr}
\tablewidth{0pc}
\scriptsize
\tablecaption{Solution for Rotation in the M87 Globular Cluster System}
\tablehead{
\colhead{Radial Range} & \colhead{N$_{\rm obs}$} & 
\colhead{$<v_r>$} & 
\colhead{$V_{rot}$}
& \colhead{Position Angle} \nl
\colhead{(arc-sec)} & \colhead{} &
\colhead{(\kms)} & \colhead{(\kms)} & \colhead{($^{\circ}$)}   }
\startdata
all & 205 & 1298 & 100 & 145 \nl
$ R < 180$ & 79 & 1331 & 112 & 162 \nl
$R > 180$ & 126 &  1275 &  95 & 134 \nl
\enddata
\end{deluxetable}

%
%
\begin{deluxetable}{cllll|llll}
\tablewidth{0pt}
\scriptsize
\tablecaption{Velocity Dispersions for the M87 Globular Cluster System}
\tablehead{
   \colhead{Radial Range}
 & \colhead{No. of Clusters}
 & \colhead{$<v_r>$}
 & \colhead{$\sigma(v_r)(N)$}
 & \colhead{$\sigma(v_r)(N-1)$\tablenotemark{a}}
 & \colhead{$<v_r>$ \tablenotemark{b}}
 & \colhead{$\sigma(v_r)(N)$ }
 & \colhead{$\sigma(v_r)(N-1)$\tablenotemark{a}}
 & \colhead{$\sigma(v_r)(N-1)$\tablenotemark{c}}
\\
   \colhead{}
 & \colhead{}
 & \colhead{}
 & \colhead{}
 & \colhead{}
 & \colhead{(derot)}
 & \colhead{(derot)}
 & \colhead{(derot)}
 & \colhead{(derot, corr)}
\\
   \colhead{(arc-sec)} 
 & \colhead{}
 & \colhead{(\kms)}
 & \colhead{(\kms)}
 & \colhead{(\kms)}
 & \colhead{}
 & \colhead{(\kms)} 
 & \colhead{(\kms)}
 & \colhead{(\kms)}
}
\startdata
$< 70$          & 19 & 1435 & 310 & 271 & 1410 & 318 & 279 (66)
\tablenotemark{d} & 260  \\
$70 < R < 103$  & 24 & 1410 & 335 & 288 & 1384 & 344 & 291 (61) & 273 \\
$105 < R < 146$ & 24 & 1220 & 290 & 273 & 1208 & 272 & 243 (51) & 221 \\
$146 < R < 197$ & 24 & 1350 & 440 & 407 & 1337 & 411 & 360 (75) & 346 \\
$198 < R < 241$ & 24 & 1235 & 327 & 298 & 1245 & 341 & 294 (61) & 276 \\
$244 < R < 295$ & 24 & 1176 & 417 & 368 & 1156 & 483 & 425 (89) & 413 \\
$296 < R < 328$ & 24 & 1271 & 366 & 339 & 1277 & 390 & 352 (73) & 337 \\
$330 < R < 382$ & 24 & 1230 & 379 & 357 & 1275 & 381 & 348 (73) & 333 \\
$386 < R < 517$ & 18 & 1443 & 556 & 500 & 1474 & 571 & 486 (118) & 476\\ 
\enddata
\tablenotetext{a}{$\sigma(v_r)$ calculated with the most discrepant cluster
omitted.}
\tablenotetext{b}{The rotation of the M87 globular cluster system has been
removed.}
\tablenotetext{c}{A correction for an instrumental uncertainty of 100 \kms~
has been removed in quadrature.}
\tablenotetext{d} {The values in parentheses are the
$1\sigma$ sampling uncertainties for these dispersions.}
\end{deluxetable}

\clearpage
\newpage
%
%
\begin{deluxetable}{rr|rr}
\tablewidth{0pc}
\scriptsize
\tablecaption{M87 $M/L$ From a Toy Model for the Mass of the M87\tablenotemark{a}}
\tablehead{
   \colhead{$R$} & \colhead{$M/L$\tablenotemark{b}} &    \colhead{$R$} & \colhead{$M/L$\tablenotemark{b}} \\
   \colhead{(kpc)} & \colhead{(solar)} & \colhead{(kpc)} & \colhead{(solar)}
 }
\startdata
  3.6 & 12 &  18.9 & 58 \\
  6.9 & 16 &  21.9 & 47 \\
  9.1 & 12 &  25.9 & 55 \\
 12.7 & 21 &  32.8 & 134 \\
 16.0 & 23 
\enddata
\tablenotetext{a}{A central point mass is used.}
\tablenotetext{b} {``Light'' is the integrated light of the
stellar component of M87 at 5500\AA.}
\end{deluxetable}

%
%
\begin{deluxetable}{rrr|rrr}
\tablewidth{0pc}
\scriptsize
\tablecaption{M87 Mass Model Inferred from its Globular Cluster System}
\tablehead{
   \colhead{$r$} & \colhead{$M(r)$} & \colhead{$<M/L>$\tablenotemark{a}} & 
   \colhead{$r$} & \colhead{$M(r)$} & \colhead{$<M/L>$}\tablenotemark{a} \\
   \colhead{(kpc)} & \colhead{(solar)} &  \colhead{(solar)}
 &  \colhead{(kpc)} & \colhead{(solar)} & \colhead{(solar)}
 }
\startdata
  3.6 & 0.7 ($\pm$0.3)x 10$^{11}$ & 4.2\tablenotemark{b} & 19.7 & 1.8 ($\pm$0.5)x 10$^{12}$ & 38.6 \\
  6.2 & 2.0 ($\pm$0.6) x 10$^{11}$ & 8.1 & 22.6 & 1.4 ($\pm$0.4) x 10$^{12}$ & 27.2 \\
  9.1 & 2.4 ($\pm$0.6) x 10$^{11}$ &  6.6 & 25.9 & 1.4 ($\pm$0.4) x 10$^{12}$ & 28.2 \\
 12.4 & 8.1 ($\pm$2.3) x 10$^{11}$  & 18.8 & 32.9 & 3.8 ($\pm$1.2) x 10$^{12}$ & 74.4 \\
 16.1 & 6.6 ($\pm$2.1) x 10$^{11}$ & 14.4 
\enddata
\tablenotetext{a}{$<M/L>$ is the mean value inside a sphere of radius $r$.}
\tablenotetext{b} {``Light'' is the integrated light of the
stellar component of M87 at 5500\AA.}
\end{deluxetable}

\clearpage
\newpage
%
%
\begin{deluxetable}{crrr}
\tablewidth{0pc}
\scriptsize
\tablecaption{The Mass of M87 from X-Ray Missions and from its Globular Cluster System}
\tablehead{
\colhead{r} & \colhead{$M(r)$\tablenotemark{a}} & 
\colhead{$M(r)$\tablenotemark{b}} & \colhead{$M(r)$\tablenotemark{c}} \nl
\colhead{(kpc)} & \colhead{(solar)} & \colhead{(solar)} & \colhead{(solar)} 
}
\startdata
Gas: \nl
44 & 6.2 x 10$^{10}$ & 7.5 x 10$^{10}$ \nl
88 & 2.1 x 10$^{11}$ & 3.0 x 10$^{11}$ \nl
 & \nl
Total Mass: \nl
44 &  6.4 x 10$^{12}$ & 1.5 x 10$^{12}$ & 3.0 x 10$^{12}$ \nl
88 & 1.3 x 10$^{13}$ & 5.0 x 10$^{12}$ \nl
\enddata
\tablenotetext{a}{Einstein data analyzed by Fabricant \& Gorenstein (1983).}
\tablenotetext{b}{ROSAT data analyzed by Nulsen \& B\"ohringer (1995).}
\tablenotetext{c}{Our analysis of the dynamics of the M87 globular cluster
system.}
\end{deluxetable}

\clearpage 
\figcaption[/scr/jlc/m87/m87_figure1.ps]
{The spatial distribution of the sample of candidate M87 globular clusters
from the Strom \et (1981) survey is shown.  Objects for which
$v_r$ were determined are indicated by filled dots.  Galactic stars
are shown as filled asterisks, while galaxies are indicated by ``G''.}
\label{fig1}
\figcaption[/scr/jlc/m87/m87_figure2.ps]
{The histogram of the heliocentric radial velocities for the
candidate M87 globular clusters.  Galaxies and M stars have been
excluded.}
\label{fig2}
\figcaption[/scr/jlc/m87/m87_figure3.ps]
{A histogram of the difference between the two independent $v_r$
determinations is shown for the 43 candidate globular clusters in M87
which have $v_r$ determined from more than one slitmask.}
\label{fig3}
\figcaption[/scr/jlc/m87/m87_figure4.ps]
{The solution for the projected rotation of M87 as inferred from the entire
sample is shown as the solid curve.  The velocities of the
individual globular clusters
in M87 are indicated as filled points.}
\label{fig4}

%
%
%

\figcaption[/scr/jlc/m87/m87_figure5.ps]
{The line-of-sight velocity dispersion
$\sigma(v_r)$ for the M87 globular cluster sample is shown
as a function of projected radius.  The open circles denote
optical measurements of $\sigma(v_r)$ for the stellar component
of M87.  Assuming constant $M/L$, the thin curve represents $\sigma(v_r)(R)$
predicted for the stellar halo of M87.  The thick curves represent $\sigma(v_r)(R)$ predicted for the globular cluster system of M87
with isotropic orbits yielding the solid line, radial orbits yielding
the dashed line, and circular orbits yielding the dot-dashed line.}
\label{fig5}

\figcaption[/scr/jlc/m87/m87_figure6.ps]
{The filled circles denote $M(r)$ as derived 
here from the globular cluster system
of M87.  The upper and lower limits on $M(r)$ from the
ROSAT data of Nulsen \& B\"ohringer (1995) are shown as the thin solid
lines.
The large open circle is the value of Mould \et (1990).
The small open circles are from Sargent \et (1978).  
The best fit power law to the globular cluster data is shown as
a solid line.}
\label{fig6}

\clearpage

\begin{figure}
\plotone{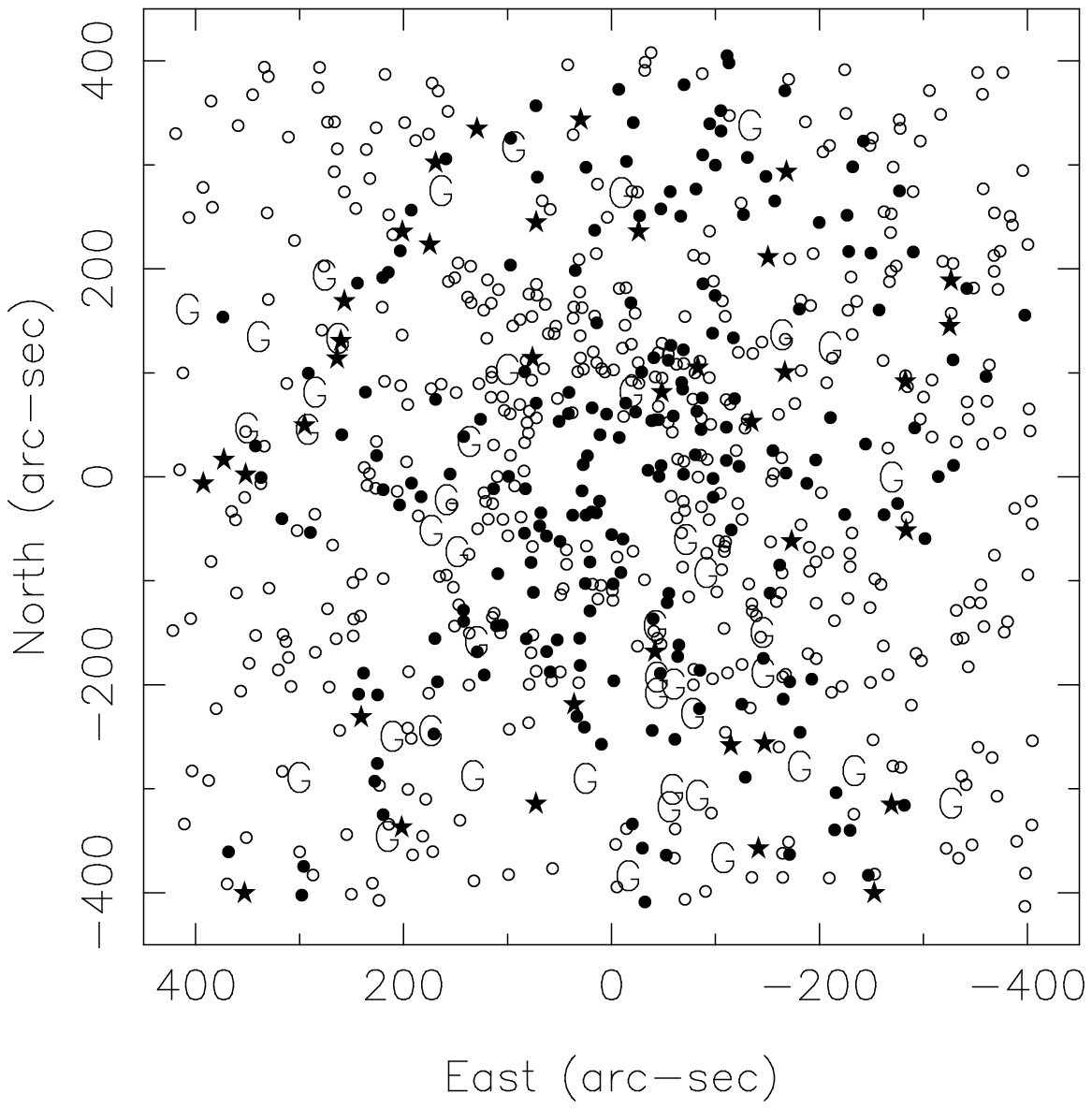}
\end{figure}

\clearpage
\begin{figure}
\plotone{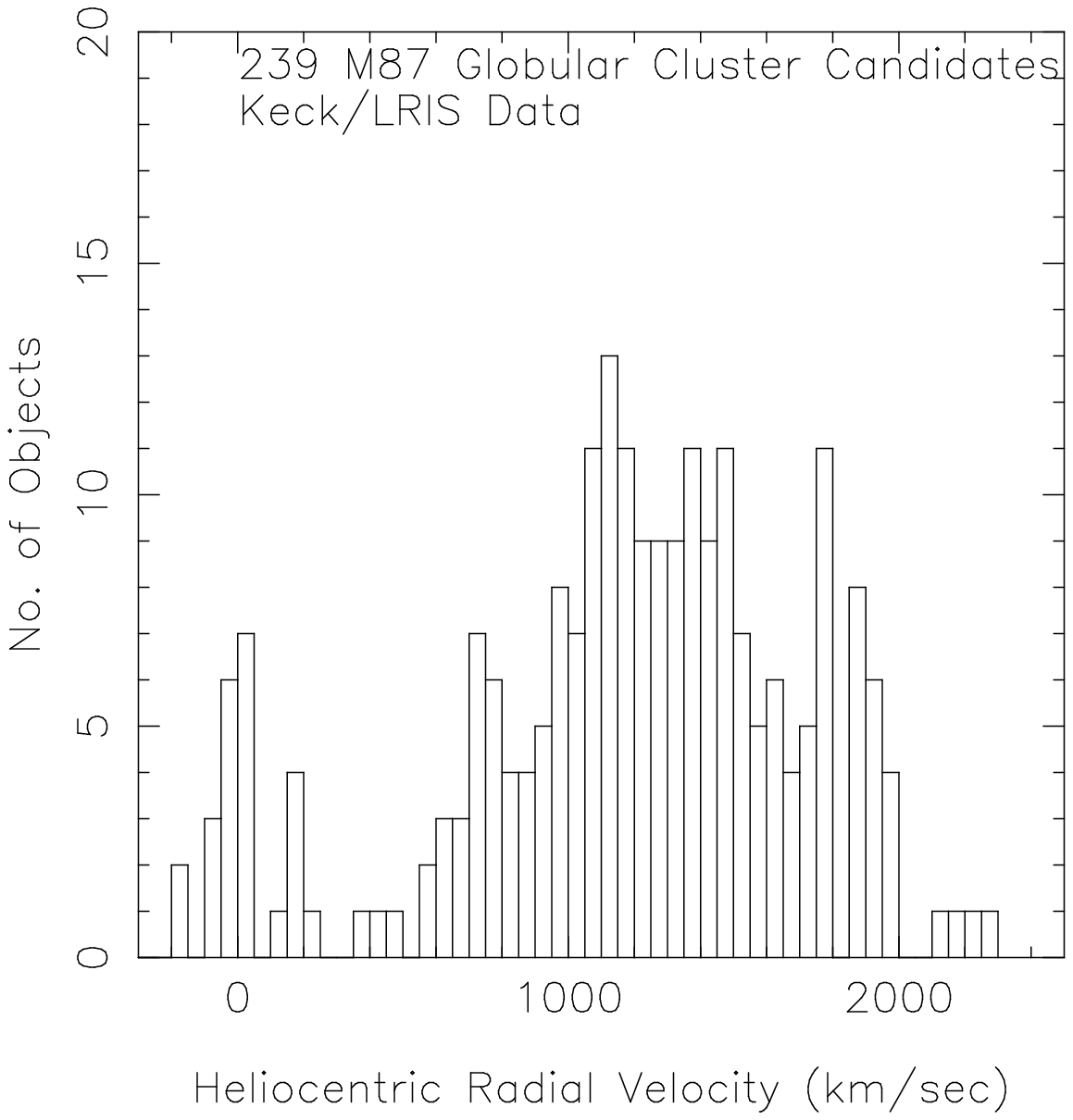}
\end{figure}

\clearpage
\begin{figure}
\plotone{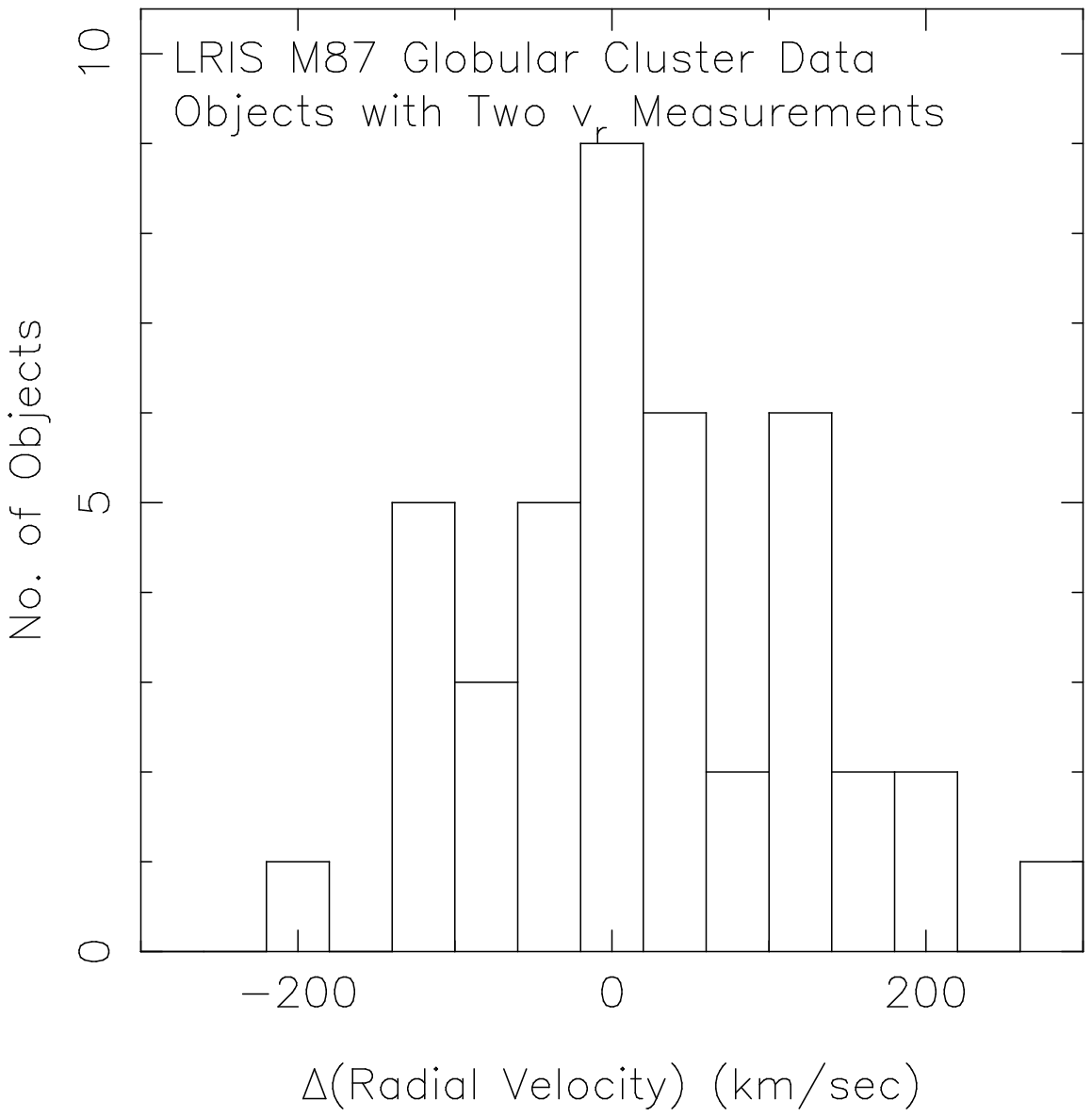}
\end{figure}

\clearpage
\begin{figure}
\plotone{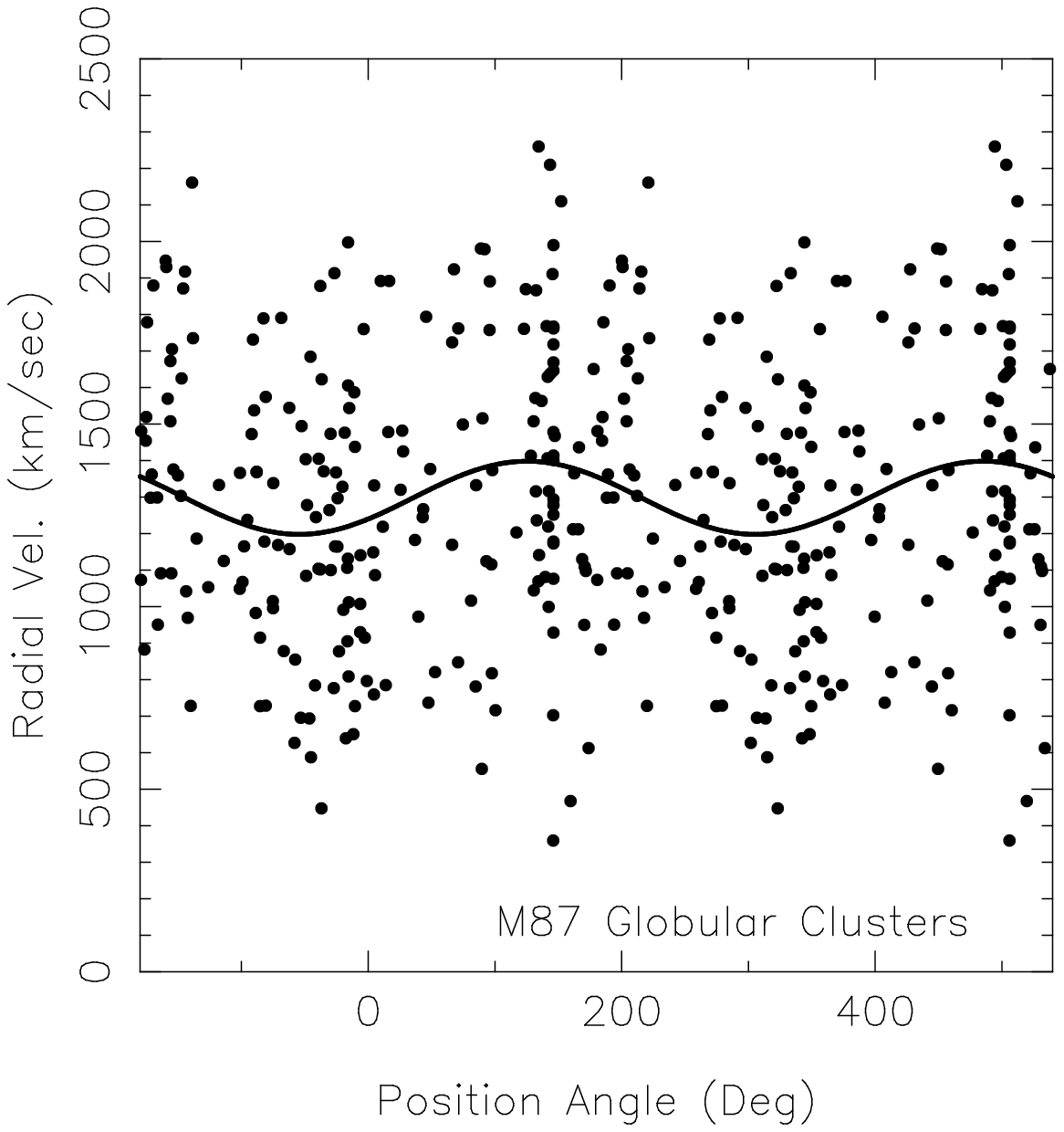}
\end{figure}

\clearpage
\begin{figure}
\plotone{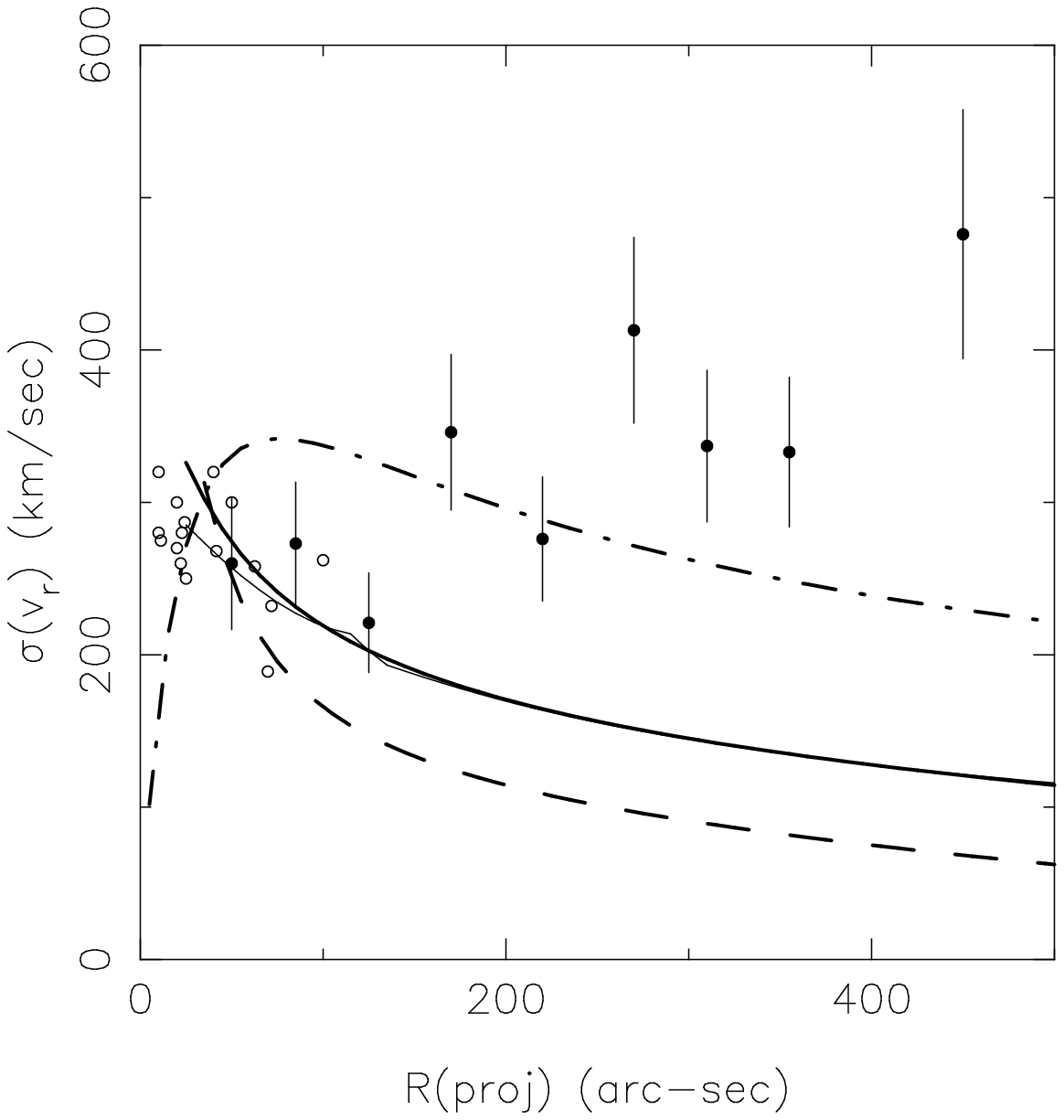}
\end{figure}

\clearpage
\begin{figure}
\plotone{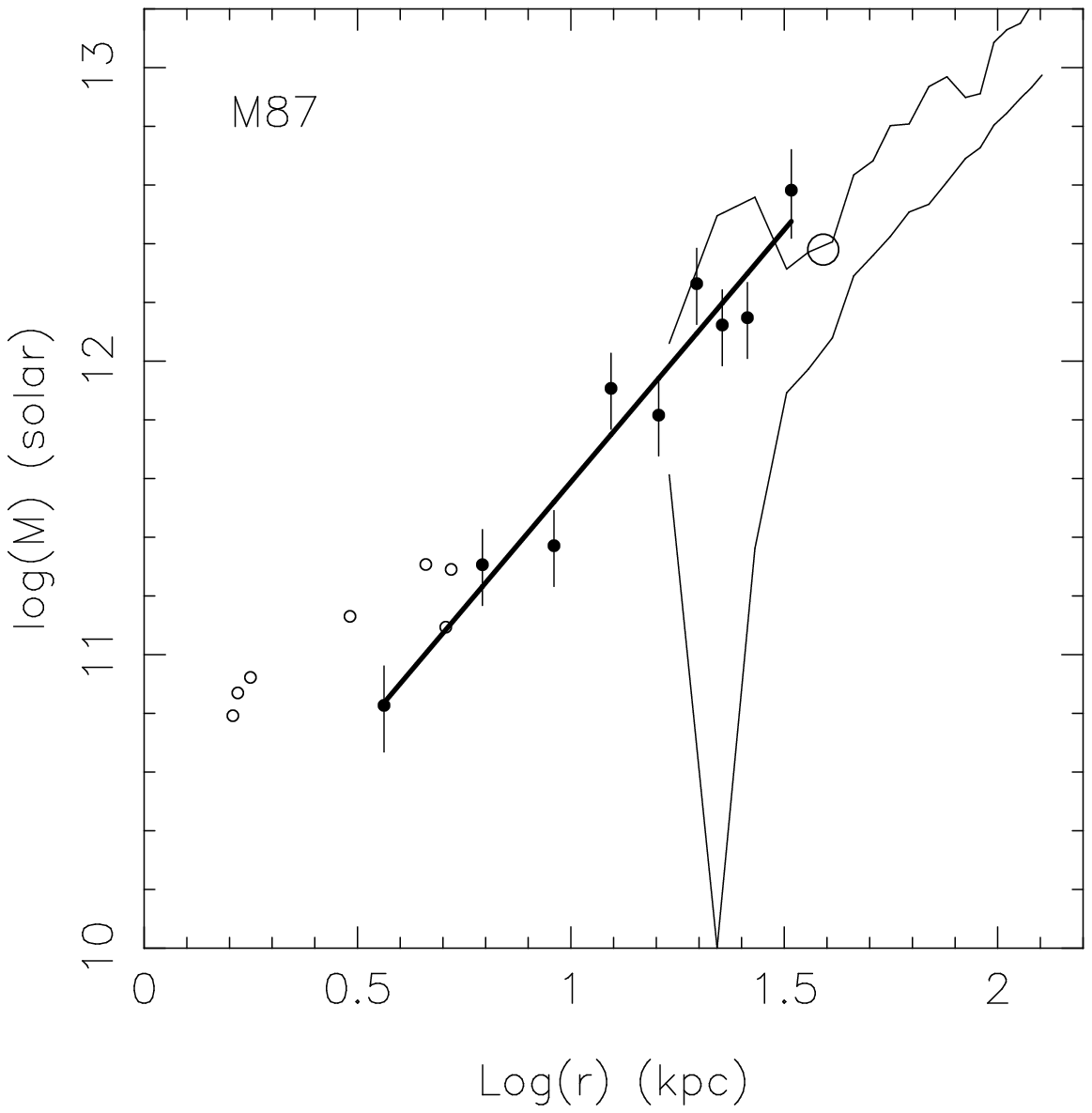}
\end{figure}

\end{document}